\documentclass[preprint,nofootinbib,aps,superscriptaddress,eqsecnum]{revtex4-2} 
\usepackage[colorlinks,citecolor=red]{hyperref}
\usepackage[usenames]{color}
\usepackage{xcolor}
\usepackage{float}
\pdfoutput=1
\usepackage{tabularx}
\usepackage{graphicx}
\usepackage{amsmath}
\usepackage{amsfonts}
\usepackage{amssymb}
\usepackage[justification=centering]{caption} 
\usepackage{caption}
\usepackage{subcaption}
\usepackage{color}
\allowdisplaybreaks
\def\bea{\begin{eqnarray}}
	\def\eea{\end{eqnarray}}
\def\be{\begin{equation}}
	\def\ee{\end{equation}}

\begin{document}
	\title{$keV$ sterile neutrino as dark matter in doublet left-right symmetric model with $A_{4}$ modular symmetry}
	\author{Ankita Kakoti}
	\email{ankitak@tezu.ernet.in}
	\affiliation{Department of Physics, Tezpur University, Tezpur 784028, India}
	
	\author{Mrinal Kumar Das}
	\email{mkdas@tezu.ernet.in}
	\affiliation{Department of Physics, Tezpur University, Tezpur 784028, India}
	\begin{abstract}
		
	Left-Right Symmetric Model(LRSM) in this work is extended with a sterile fermion per generation, the lightest of the same is considered to be a suitable dark matter candidate for the study and analysis of the associated properties. The model has been realized using $A_{4}$ modular symmetry, the advantage being the non-requirement of the use of extra fields, hence keeping the model minimal. Because of the extension of LRSM with the sterile fermion, the neutrino mass in this work will be generated by the double seesaw mechanism as described thoroughly within the manuscript. And for phenomenological studies, we have considered neutrinoless double beta decay the details of which have been discussed thoroughly within the work.

	\end{abstract}
	\maketitle
	\newpage
\section{\label{lrsm1}\textbf{Introduction}}
One of the most intriguing and open questions in the area of particle physics and cosmology is the existence of dark matter and its properties. Dark matter was first proposed by Fritz Zwicky in the year 1933 while studying Coma cluster of galaxies. The analysis of observational data of the satellite-based experiments, Wilkinson Microwave Anisotropy Probe (WMAP) \cite{WMAP:2012fli,WMAP:2006bqn} and Planck \cite{Planck:2018vyg} that look for very tiny anisotropies in CMBR suggest that the total energy density of the Universe contains $69\%$ dark energy, $27\%$ dark matter and $4\%$ ordinary matter. But, unfortunately Standard Model (SM) could not shed any light upon the explanations regarding the same. There are several beyond SM frameworks which take into account the study of dark matter and associated phenomenology.\\
A BSM framework which can incorporate within itself explanations regarding the existence of dark matter and at the same time provide a suitable platform for analysis of phenomenology like neutrinoless double beta decay, lepton flavor violation, leptogenesis etc. is Left-Right Symmetric Model (LRSM). A recent work \cite{Dev:2025fcv} has suggested that the neutral component of scalar triplet $\Delta_{R}$ can be a suitable dark matter candidate, but in this work, we have extended LRSM with a sterile fermion per generation and the lightest of the same is considered as the dark matter candidate. The experiments like atmospheric, solar and long baseline experiments are well-fitted within the regime of three neutrino mixing, but some current experiments like LSND\cite{LSND:1997vun}, MiniBooNE\cite{MiniBooNE:2020pnu} etc. suggest the existence of fourth massive neutrino usually called 'sterile'. Also, anomalies in SAGE\cite{SAGE:1998fvr} and GALLEX\cite{Naumov:2019kwm} have provided a hint towards the existence of the fourth state of neutrinos. A natural way to produce sterile neutrinos is by their admixture to the active neutrinos and the only way to reveal the existence of sterile neutrinos in terrestrial experiments is through the effects generated by their mixing with the active neutrinos. According to the Planck data, about $26.8\%$ of the energy density in the Universe comprises of dark matter (DM). The present dark matter abundance is \cite{Planck:2015fie},
\begin{equation*}
	\Omega_{DM}h^{2} = 0.1187 \pm 0.0017
\end{equation*}
Incorporating the sterile fermion per generation, the light neutrino mass within this framework will now be determined by the double seesaw mechanism, that is we will be working on LRSM with double Seesaw. After that we calculate the relic abundance and also decay rate of the corresponding DM candidate and also analyze the results using Lyman-$\alpha$\cite{Baur:2017stq} and X-ray constraints \cite{Ng:2019gch}.\\
In the current work, we also analyze the impact of incorporating double seesaw mechanism in LRSM on phenomenology like neutrinoless double beta decay ($0\nu\beta\beta$). $0\nu\beta\beta$ is a lepton number violating process which if proven by experiments will mean that neutrinos are Majorana in nature. For study of the phenomenology, we are mainly focussed on the calculation of the effective neutrino mass as well as half-life of the decay \cite{Ge:2017erv}. The experiments like KamLAND-Zen\cite{Shirai:2013jwa} and GERDA\cite{GERDA:2020xhi} which uses Xenon-136 and Germanium-76 respectively have improved the lower bound on the half-life of the decay process. However, KamLAND-Zen imposes the best lower limit on the half life as $T_{1/2}^{0\nu}>1.07\times10^{26}$ yr at 90 percent confidence level (CL) and the corresponding upper limit of the effective Majorana mass in the range (0.061-0.165)eV. As studied in our previous works like \cite{Kakoti:2023isn,Kakoti:2023xkn}, in LRSM there are eight contributions of $0\nu\beta\beta$ corresponding to the presence of several new particles in the model. However, in the current work as will be discussed in the succeeding sections we will be working with doublet LRSM and the contributions that we take into account are standard light neutrino, heavy right-handed as well as sterile neutrino contributions.Although LRSM stands as a suitable platform for the study of lepton flavor violation(LFV), but as we will be considering doublet LRSM, the LFV processes will arise only at the loop level, and hence the study of LFV within the context of doublet LRSM with double seesaw mechanism has been kept for future work.\\
In the present work, we have used the global $3\sigma$ \cite{Esteban:2024eli} values for the calculation of the relevant neutrino parameters. Organization of this paper follows, section \ref{lrsm2}, where we discuss the model and the origin of neutrino mass. Section \ref{lrsm3} describes the realization of the model with $A_{4}$ modular symmetry and also the incorporation of double seesaw mechanism within the model. In section \ref{lrsm4}, we have discussed the presence of dark matter in extended left-right symmetric model and section \ref{lrsm41} theoretically discusses $0\nu\beta\beta$ within the model. Section \ref{lrsm5} presents the numerical analyses and results of the work and in section \ref{lrsm6} and \ref{lrsm7} we have presented the discussion and conclusion of the present work respectively.\\
Although sterile neutrino dark matter is not completely a new concept in the context of left-right symmetric model, but the most attractive feature of the current work lies in the augmentation of modular symmetry with the model and the better elucidation of constraints obtained for parameters like decay rate as well as mass of dark matter candidate, as has been illustrated in the succeeding sections.

\section{Doublet Left-Right Symmetric Model and Neutrino Mass.}\label{lrsm2}

Left-Right Symmetric Model is described by the gauge group $SU(3)_C \otimes SU(2)_L \otimes SU(2)_R \otimes U(1)_{B-L}$, from which it can be understood that unlike SM, both the left-handed and right handed particles are taken as doublets in the said model. The scalar sector of LRSM can consist of the following fields,\\
\begin{itemize}
	\item one Higgs bidoublet $\phi(1,2,2,0)$ and two scalar triplets $\Delta_L(1,3,1,2)$ and $\Delta_R(1,1,3,2)$ \cite{Maiezza:2016ybz,Senjanovic:2016bya}.
	\item one Higgs bidoublet $\phi(1,2,2,0)$ and two scalar doublets $H_{L}(1,2,1,-1)$ and $H_{R}(1,1,2,-1)$ \cite{Senjanovic:1978ev,Mohapatra:1977be,Borah:2025fkd}.
\end{itemize}

In the current work, we have used the scalar sector consisting of one Higgs bidoublet and two scalar doublets, and this case is known as the doublet left-right symmetric model (DLRSM). In addition to triplet left-right symmetric model (TLRSM) and DLRSM, there can also be other realizations of the model that have experimental consequences \cite{Das:2016akd,Das:2017hmg}.\\
Study of dark matter in LRSM has been done in previous works like \cite{Nemevsek:2012cd,Dev:2025fcv,Boruah:2022csq}, however, in this work we extend LRSM with a sterile fermion per generation represented by $S_{L}(1,1,1,0) $ under the gauge group describing the model. The lightest of the sterile fermion acts as a suitable dark matter candidate for the model under consideration. The resulting neutrino mass will be generated by the double seesaw mechanism, discussed as under \cite{Patel:2023voj}. Here, the fermion sector will consist of the usual quarks, leptons and the sterile fermion. The scalar sector will consist of one Higgs bidoublet and two scalar doublets, as shown below,\\
\textbf{Fermion sector}
\begin{equation}
	q_{L} = \begin{pmatrix}
		u_{L}\\
		d_{L}
	\end{pmatrix} \sim (3,2,1,1/3),
	q_{R} = \begin{pmatrix}
		u_{R}\\
		d_{R}
	\end{pmatrix} \sim (3,1,2,1/3)
\end{equation}
\begin{equation}
	l_{L} = \begin{pmatrix}
		\nu_{L}\\
		e_{L}
	\end{pmatrix} \sim (1,2,1,-1),
	l_{R} = \begin{pmatrix}
		\nu_{R}\\
		e_{R}
	\end{pmatrix} \sim (1,1,2,-1)
\end{equation}
\textbf{Higgs sector}
\begin{equation}
	H_{L} = \begin{pmatrix}
		h_{L}^{+}\\
		h_{L}^{0}
	\end{pmatrix} \sim (1,2,1,-1),
	H_{R} = \begin{pmatrix}
		h_{R}^{+}\\
		h_{R}^{0}
	\end{pmatrix} \sim (1,1,2,-1),
	\phi = \begin{pmatrix}
		\phi_{1}^{0} & \phi_{2}^{+}\\
		\phi_{1}^{+} & \phi_{2}^{0} 
	\end{pmatrix}\sim (1,2,2,0)
\end{equation}
The Yukawa Lagrangian in case of the double seesaw mechanism is given as shown in equation \eqref{e1}\cite{Patel:2023voj}.
\begin{equation}
	\label{e1}
	\mathcal{L_Y}= \overline{l_L}\phi{l_R}Y_{ij}+\overline{l_L}\tilde{\phi}{l_R}\tilde{Y_{ij}}+F(\overline{l_R})H_{R}\overline{S_{L}}+\mu \overline{ S_{L}^{C}}S_{L} +h.c
\end{equation}
\begin{equation}
	\supset M_{D}\overline{\nu_{L}}\nu_{R}+M \overline{\nu_{R}}S_{L}+\mu \overline{ S_{L}^{C}}S_{L}+h.c
\end{equation}
The spontaneous symmetry breaking (SSB) in the model takes place in two steps. The LRSM gauge group breaks into the SM gauge group by the VEV of scalar doublet $H_{R}$ and finally it breaks down into $U(1)_{(em)}$ group by VEV of the Higgs bidoublet $\phi$. It is to be noted here that the presence of $H_{L}$ is only for the left-right invariance and $<H_{L}>=0$. Zero VEV of $H_{L}$ ensures that it does not take part in spontaneous symmetry breaking process, as well as do not introduce any extra terms in the Lagrangian.\\
After SSB, the $9\times9$ neutrino mass matrix is depicted as,
\begin{equation}{\label{ex1}}
	M_{\nu}^{9\times9}=\begin{pmatrix}
		0 & M_{D} & 0 \\
		M_{D}^{T} & 0 & M\\
		0 & M^{T} & \mu
	\end{pmatrix}
\end{equation}
where, $M_{D}$ represents the Dirac mass matrix obtained from the term connecting $\nu_{L}$ and $\nu_{R}$. $M$ here represents the mixing matrix obtained as a result of mixing between the right-handed neutrinos and newly introduced sterile fermions, and $\mu$ is the Majorana mass matrix depicting the mass term for the sterile neutrinos, which are the newly added fermions in the model. It is to be noted here, that unlike the Standard Model of particle particles, the right-handed neutrinos in LRSM are considered to be active as they interact with their left handed counterparts and also with the additional fermions.
Following the mass hierarchy $M_{D}$, $M$ $>>$ $\mu$, 
\begin{equation}
	\label{e2}
	m_{\nu} = M_{D}(M^{T})^{-1}\mu M^{-1}M_{D}^{T}
\end{equation}
The double seesaw mechanism is performed by incorporating the type-I seesaw mechanism twice. Now, considering $M_{R}$ as the Majorana mass matrix corresponding to the heavy right-handed neutrinos in the model, for the fact that $M_{R}$ $>>$ $M_{D}$, we get the mass of the heavy neutrinos in the form \cite{Patra:2023ltl},
\begin{equation}
	\label{ee1}
	M_{R} = M.\mu^{-1}M^{T}
\end{equation}
which is the implication of the first seesaw. The second seesaw shows that the resulting neutrino mass matrix becomes linearly dependent on the sterile neutrino mass $\mu$ as shown in expression \eqref{e2}. However, as is also evident from the said equation that the light neutrino mass is doubly suppressed by the active-sterile mixing matrix and hence the named has been dubbed as, 'double seesaw mechanism' \cite{Grimus:2013tva}.

\section{\label{lrsm3}Doublet LRSM using modular group of level 3 ($\Gamma(3)$) and $k_{Y}=2$ .}
Left-Right Symmetric Model with discrete flavor symmetries has been studied in several earlier works like \cite{Duka:1999uc,Boruah:2022bvf,Sahu:2020tqe,Rodejohann:2015hka}. However, advent of the concept of modular symmetry \cite{Feruglio:2017spp,King:2020qaj} put forward a way to study generation of neutrino masses within a model without the use of extra particles or flavons. In this work we have realized LRSM with $A_{4}$ modular symmetry where the model contains usual particle content, without the use of any flavons \cite{Sahu:2020tqe}. The charge as well as weight assignments for the particles under $A_4$ \cite{Abbas:2020qzc} have been shown in table \ref{tab:Table 1} where $k$ represents the modular weight of each particle. Now, while assigning modular weights it is to be noted that sum of the modular weights in each term of the Lagrangian should add up to be zero, and also in the scenario that we are working, modular forms cannot be assigned negative weights. A brief discussion of modular symmetry has been presented in the appendix of the manuscript.\\
The Yukawa Lagrangian for the leptonic sector in LRSM is given by Eq.\eqref{e1} from which we can write the Yukawa Lagrangian of our $A_4$ modular symmetric LRSM, for the lepton sector, by introducing Yukawa coupling in the form of modular forms $Y$ given as,
\begin{equation}
	\label{e3}
	\mathcal{L_Y}= \overline{l_L}\phi{l_R}Y+\overline{l_L}\tilde{\phi}{l_R}Y+fY(\overline{l_R})H_{R}\overline{S_{L}}+ \mu \overline{ S_{L}^{C}}S_{L}
\end{equation}
The fermions transforming as triplets under the $A_{4}$ discrete symmetry group ensures in keeping the Lagrangian invariant under the said gauge group because the modular form $Y$ is also a triplet and the scalars in the model transform as singlets under $A_{4}$ group. In determining so, we have used the multiplication rules of $A_{4}$ group as stated in the appendix of the manuscript. In particular, the standard $A_{4}$ tensor product rules yield,
\begin{equation*}
	3 \otimes 3 = 1 \oplus 1' \oplus 1'' \oplus 3_A \oplus 3_S
\end{equation*}
As such, the product of three triplets can be contracted sequentially to give us an $A_{4}$ singlet. If we take an example of three triplet fields say, $a$,$b$ and $c$, so first the product of $a$ and $b$ can form the symmetric (or antisymmetric) triplet combination of $A_{4}$ group and then can contract with the third triplet $c$ to obtain the overall singlet combination.\\
\begin{table}[!h]
	\caption{\centering Charge and modular weight assignments of particle content of the model.}
	\label{tab:Table 1}
	\centering
	\begin{tabular}{|c|c|c|c|c|c|c|}
		\hline
		Gauge group & $l_L$ & $l_R$ & $\phi$  & $H_R$ & $S_{L} $\\
		\hline
		$SU(3)_C$ & 1 & 1 & 1 & 1 & 1\\
		\hline
		$SU(2)_L$ & 2 & 1 & 2  & 1 & 1 \\
		\hline
		$SU(2)_R$ & 1 & 2 & 2 &  2 & 1 \\
		\hline
		$U(1)_{B-L}$ & -1 & -1 & 0 & -1 & 0 \\
		\hline
		$A_4$ & 3 & 3 & 1 & 1 & 3\\
		\hline 
		$k$ & 2 & 2 & -2 & -1 & -1 \\
		\hline
	\end{tabular}
\end{table}
\begin{table}[!h]
	\caption{\centering Charge assignment and modular weight for the corresponding modular Yukawa form for the model.}
	\label{tab:Table 2}
	\centering
	\begin{tabular}{|c|c|}
		\hline
		& Y (modular forms)\\
		\hline
		$A_4$ & $3$ \\
		\hline
		$k_Y$ & $2$ \\
		\hline
	\end{tabular}
\end{table}
The Yukawa couplings $Y = (Y_1,Y_2,Y_3)$ are expressed as modular forms of level 3. Table \ref{tab:Table 2} shows the charge assignment for the modular form of level 3 and its corresponding modular weight. To determine the resulting light neutrino mass, we need to first determine the associated mass matrices, so using the $A_{4}$ multiplication rules we get the Dirac mass matrix in terms of the modular forms $(Y_1,Y_2,Y_3)$ ,
\begin{equation}
	\label{e4}
	M_D =v\begin{pmatrix}
		2Y_1 & -Y_3 & -Y_2\\
		-Y_2 & -Y_1 & 2Y_3\\
		-Y_3 & 2Y_2 & -Y_1
	\end{pmatrix}
\end{equation}
where, $v$ is VEV of the Higgs bidoublet $\phi$ and $v=\sqrt{k^{2}+k'^{2}} \simeq 246 GeV$, $k$ and $k'$ representing the VEVs of the neutral components of $\phi$. The active-sterile neutrino matrix is given by,
\begin{equation}
	\label{e:a}
	M =v_R f\begin{pmatrix}
		2Y_1 & -Y_3 & -Y_2\\
		-Y_3 & 2Y_2 & -Y_1\\
		-Y_2 & -Y_1 & 2Y_3
	\end{pmatrix}
\end{equation}
where, $v_R$ is the VEV for the scalar doublet $H_R$ and $f$ is a dimensionless arbitrary parameter. The Majorana sterile neutrino mass matrix is given by,
\begin{equation}
	\label{e:b}
	\mu = s \begin{pmatrix}
		2Y_1 & -Y_3 & -Y_2\\
		-Y_3 & 2Y_2 & -Y_1\\
		-Y_2 & -Y_1 & 2Y_3
	\end{pmatrix}
\end{equation}
$s$ being an arbitrary parameter with dimensions of $keV$.\\
The light neutrino mass can be determined using the expression \eqref{e2}, which will be given in the current work as,
\begin{equation}
	\label{e5}
	M_\nu = \begin{pmatrix}
		\frac{2 s v^2 Y_1}{f^2 {v_{R}}^2} & \frac{- s v^2 Y_2}{ f^2 {v_{R}}^{2} }& -\frac{s v^2 Y_3}{f^2 {v_{R}}^2} \\
		\frac{- s v^2 Y_2}{ f^2 {v_{R}}^{2}} & \frac{2 s v^2 Y_3}{f^2 {v_{R}}^{2}} & -\frac{sv^2 Y_1}{f^2 {v_{R}}^2} \\
		-\frac{ s v^2 Y_3}{f^2 {v_{R}}^2} & -\frac{sv^2 Y_1}{f^2 {v_{R}}^2} & \frac{2 s v^2 Y_2}{f^2 {v_{R}}^2}
	\end{pmatrix}
\end{equation}
It is evident from \eqref{e5} that $M_{\nu}$ is symmetric in  nature. As we have determined the light neutrino mass, we will now move to the analysis of different phenomenology including dark matter and $0\nu\beta\beta$ as shown in the succeeding sections.

\section {\label{lrsm4}Dark matter in Left-Right Symmetric Model}
Dark Matter (DM) and its existence in the Universe is one of the most intriguing open questions in the area of cosmology and particle physics. The properties of dark matter as stated in \cite{Taoso:2007qk,Bergstrom:2009ib} are quite complex to take into account considering the search for a perfect DM candidate in the existing enumeration of different mass models. However, several studies are going on taking into account different fields that are assumed to act as a suitable DM candidate \cite{Gautam:2019pce,Boruah:2021ktk,Sarma:2022qka,Sarma:2021icl}. Depending upon the model taken into consideration, the fields can be fermion or scalar fields respectively. \\
In this work, extending LRSM with a sterile fermion per generation we consider the lightest of the same to be a suitable warm dark matter (WDM) candidate \cite{Merle:2017jfn,Drewes:2016upu}. Here, neutrino mass is generated by double seesaw mechanism and this mechanism has the advantage of allowing for a non-zero mixing between the active and additional sterile states. One of the most important criteria for being a DM candidate is that it has to be stable, at least on the cosmological scale. However, the lightest of the sterile neutrino is not perfectly stable, it can decay radiatively to a photon and an active neutrino via the process $S \rightarrow \nu + \gamma$, which produces an X-ray signal. However, as stated in many literature \cite{Drewes:2016upu}, the decay rate is negligible owing to the small mixing between active and sterile states. Now, since the sterile neutrino is a fermionic dark matter candidate, lower bound exists on its mass known as the Tremaine-Gunn bound \cite{Boubekeur:2023fqo}. An upper limit on the dark matter mass is also obtained from the X-ray constraints and as a matter of fact, direct and indirect detection of DM also imposes significant constraints on the sterile neutrino as a DM candidate.\\
Any stable neutrino state having non-zero mixing with the active neutrino state is produced through the active-sterile neutrino conversion. Thus, the DM abundance is produced through the mixing of the active-sterile neutrino states. The mechanism of non-resonant production of sterile neutrino is known as the Dodelson-Widrow(DW) mechanism \cite{Merle:2015vzu}.\\
To calculate the relic abundance as well as decay rate of the dark matter candidate, we have used formulae stated in equations \eqref{e7} and \eqref{e8}.
The resulting relic abundance can be given by the expression \cite{Asaka:2006nq},\\
\begin{equation}
	\label{e6}
	\Omega_{DM}h^{2} = 1.1 \times 10^{7} \Sigma C_{\alpha} (m_{s})|U_{\alpha s}|^{2}\Bigg(\frac{m_{s}}{keV}\Bigg)^{2}, \alpha = e , \mu , \tau
\end{equation}
where $C_{\alpha}(m_s)$ represents active flavor dependent coefficients, which can numerically be computed by solving Boltzmann equations.
Equation \eqref{e6} can be simplified to,
\begin{equation}
	\label{e7}
	\Omega_{DM}h^{2} \approxeq 0.3\Bigg(\frac{sin^{2} 2\theta_{DM}}{10^{-10}}\Bigg)\Bigg(\frac{m_{s}}{100keV}\Bigg)^{2}
\end{equation}
where, $sin^{2}\theta_{DM} = 4\sum_{\alpha = e,\mu,\tau}|U_{\alpha s}|^{2}$, $|U_{\alpha s}|$ being the active-sterile neutrino mixing and $m_{s}$ is the mass of the lightest sterile neutrino which is the DM candidate.\\
The decay rate is given by \cite{Gautam:2019pce},
\begin{equation}
	\label{e8}
	\Gamma = 1.38 \times 10^{-32}\Bigg(\frac{sin^{2} 2\theta_{DM}}{10^{-10}}\Bigg)\Bigg(\frac{m_{s}}{100keV}\Bigg)^{5} s^{-1}
\end{equation}
From  equations \eqref{e7} and \eqref{e8} it is evident that the decay rate and the relic abundance both depend upon the mixing of the active-sterile states and mass of the DM candidate. Hence, the same set of model parameters that can produce correct neutrino phenomenology can also be used to evaluate the relic abundance and the decay rate of the sterile neutrino.

\section{\label{lrsm41}Neutrinoless Double Beta Decay}

As already stated earlier that there are eight contributions of neutrinoless double beta decay in the context of LRSM, however,
in the current scenario, we are working with a very high scale for the mass of right-handed gauge boson $(W_{R})$, so as to keep the neutrino mass small. And as such, the contribution to $0\nu\beta\beta$ due to purely right handed current gets suppressed as the contribution from purely right handed current is proportional to $\frac{1}{M_{W_R}^{4}}$. Similarly, the momentum dependent contributions that is, $\lambda$ and $\eta$ contributions are also suppressed as they are dependent on the terms $\frac{1}{M_{W_R}^2}$ and $\tan \theta_{LR}$, $\theta_{LR}$ being the mixing angle corresponding to left and right gauge bosons ($W_{L}$ and $W_{R}$) and it is small in the framework under consideration. Hence, the leading order contributions to $0\nu\beta\beta$ in the current work are the ones arising from purely left-handed current due to the exchange of light neutrino, heavy right-handed Majorana neutrino and sterile neutrino.\\
The $9\times 9$ neutrino mass matrix is expressed as in equation \eqref{ex1},
\begin{equation*}
	M_{\nu}^{9\times9}=\begin{pmatrix}
		0 & M_{D} & 0 \\
		M_{D}^{T} & 0 & M\\
		0 & M^{T} & \mu
	\end{pmatrix}
\end{equation*}
The diagonalization of $M_{\nu}$ leads to the following relation between the fields of the neutral fermions in the flavor basis and in the mass eigenstate basis \cite{Patra:2023ltl},
\begin{equation}
	\begin{pmatrix}
		\nu_{\alpha L}\\
		{\nu^{C}_{R}}_{\beta L}\\
		S_{\gamma L}
	\end{pmatrix}=\begin{pmatrix}
		V^{\nu\nu}_{\alpha i}&V^{\nu\nu_{R}}_{\alpha j}&V^{\nu S}_{\alpha k}\\
		V^{\nu_{R}\nu}_{\beta i}&V^{\nu_{R}\nu_{R}}_{\beta j}&V^{\nu_{R}S}_{\beta k}\\
		V^{S\nu}_{\gamma i}&V^{S\nu_{R}}_{\gamma j}&V^{SS}_{\gamma k}
	\end{pmatrix}\begin{pmatrix}
		\nu_{iL}\\
		{\nu^{C}_{R}}_{j L}\\
		S_{kL}
	\end{pmatrix}
\end{equation}
where, $C$ is the charge conjugation operator, the indices $\alpha$,$\beta$ and $\gamma$ run over the three generations of light left-handed neutrinos, heavy right-handed and sterile neutrinos in the flavor basis respectively, and the indices $i$,$j$ and $k$ run over corresponding mass eigenstates.\\
For dominant contributions of neutrinoless double beta decay in our model, the lepton Lagrangian can be given as \cite{Patra:2023ltl},
\begin{equation}
	\mathcal{L}^{l}_{CC} = \frac{g_{L}}{\sqrt{2}}[\overline{e_{L}}\gamma_{\mu}{(V^{\nu\nu}_{ei}\nu_{i})}W^{\mu}_{L}]+h.c+\frac{g_{R}}{\sqrt{2}}[\overline{e_{R}}\gamma_{\mu}(V^{\nu_{R}\nu_{R}}_{ej}{\nu_{R}}_{j}+V^{\nu_{R}S}_{ek}S_{k})W^{\mu}_{R}]+h.c
\end{equation}
That is, the dominant contributions come from 
\begin{itemize}
	\item the standard mechanism due to the exchange of light neutrinos $\nu_{i}$, mediated by $W_{L}$.
	\begin{figure}[!h]
		\centering
		\includegraphics[width=0.5\linewidth]{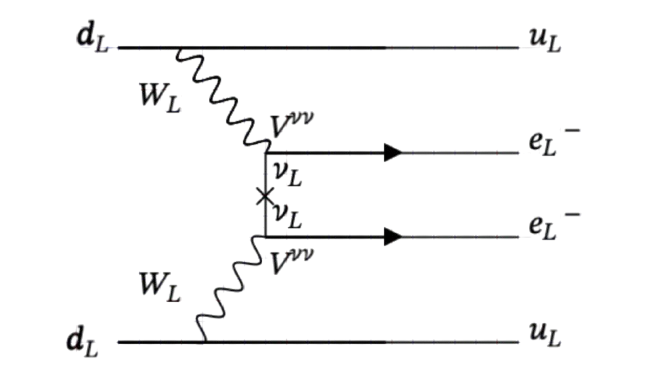}
		\caption{\centering Feynmann diagram for $0\nu\beta\beta$ mediated by the exchange of standard light neutrinos, $\nu_{L}$.}
		\label{fd1}
	\end{figure}
	\item new contributions due to the exchange of heavy neutrinos ${\nu_{R}}_{1,2,3}$ and sterile neutrinos $S_{1,2,3}$ mediated by the gauge boson $W_{R}$.
	\begin{figure}[!h]
		\centering
		\includegraphics[width=0.38\linewidth]{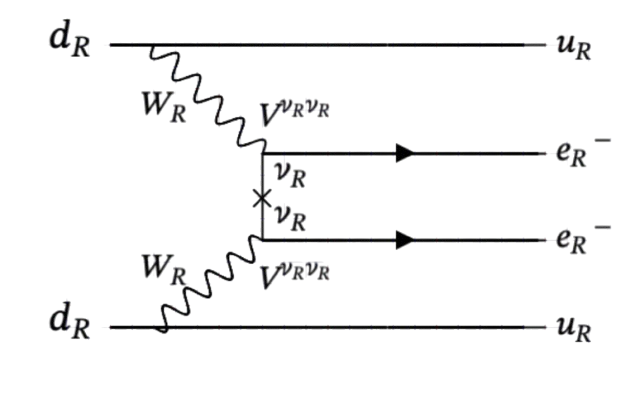}
		\includegraphics[width=0.38\linewidth]{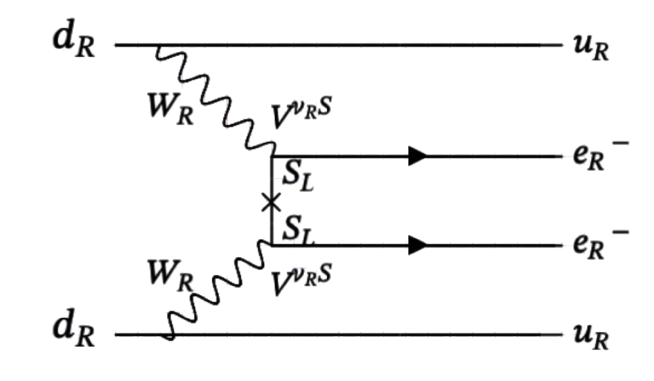}
		\caption{\centering Feynmann diagram for $0\nu\beta\beta$ mediated by the exchange of heavy RH neutrinos(left) and sterile neutrinos(right).}
		\label{fd2}
	\end{figure}
\end{itemize}
\section{\label{lrsm5}Numerical Analysis and Results}

\begin{figure}[!h]
	\includegraphics[scale=0.4]{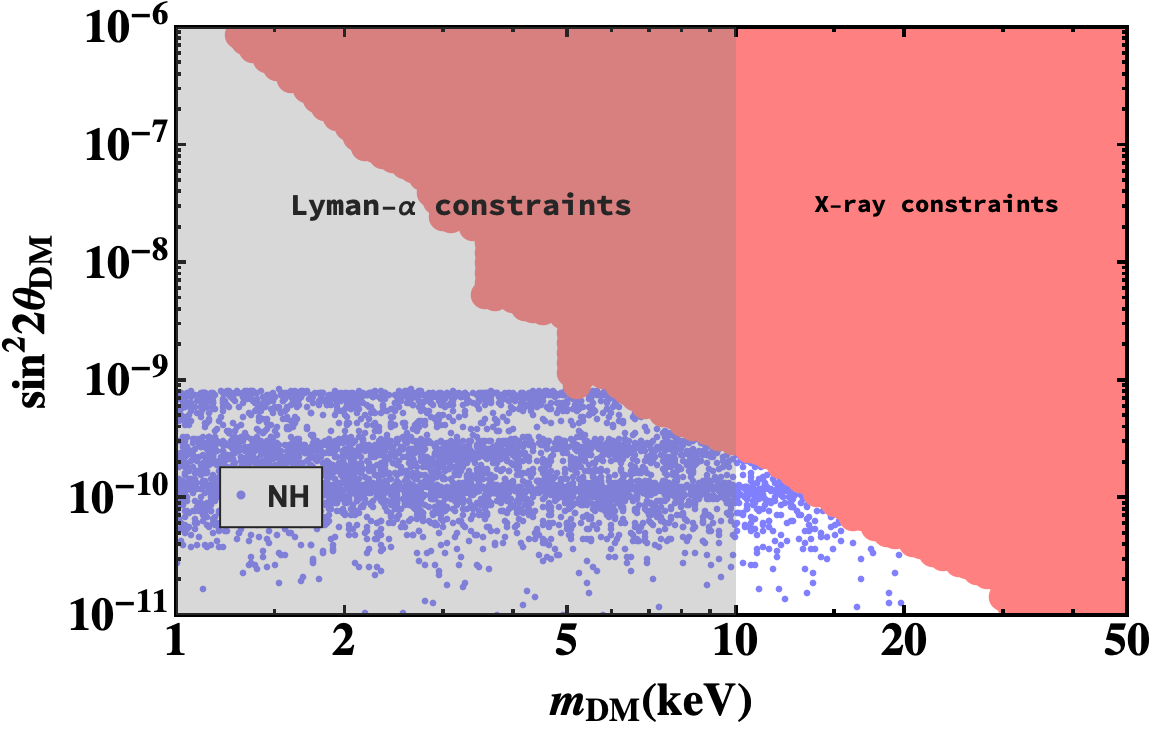}
	\includegraphics[scale=0.4]{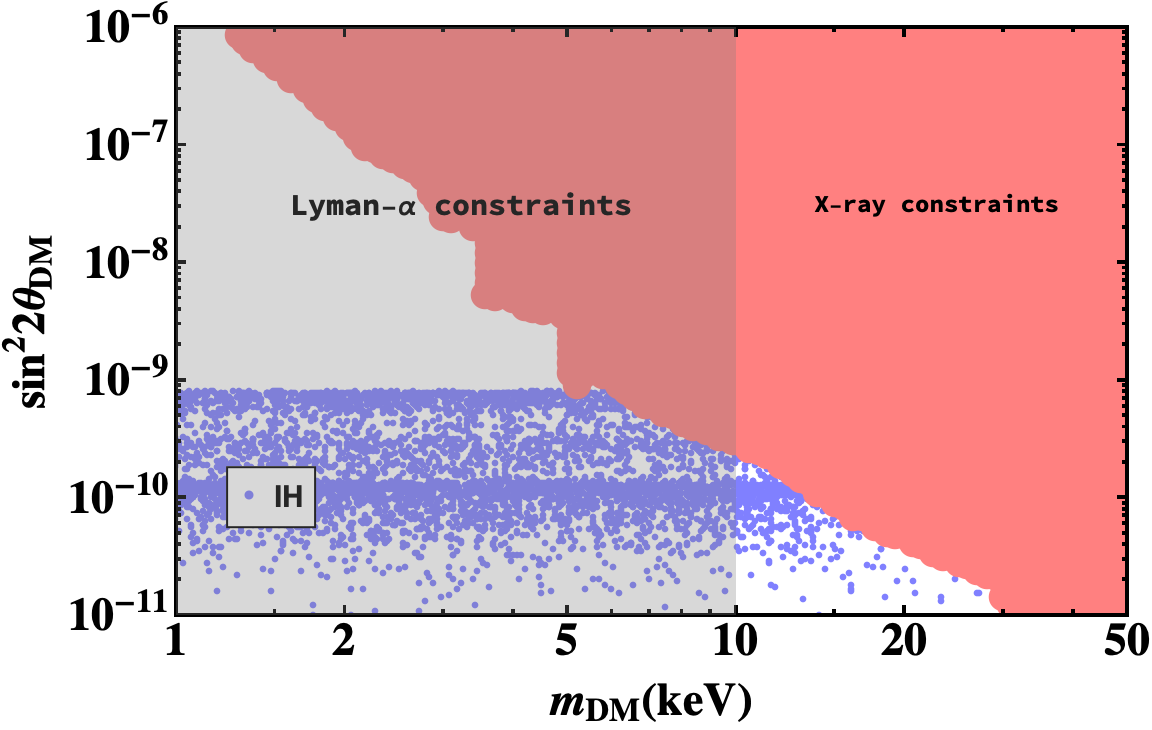}
	\caption{\centering \label{f4}Variation of mass of DM candidate with active-sterile mixing angle for NH(left) and IH(right).}
\end{figure}

\begin{figure}[!h]
	\includegraphics[scale=0.4]{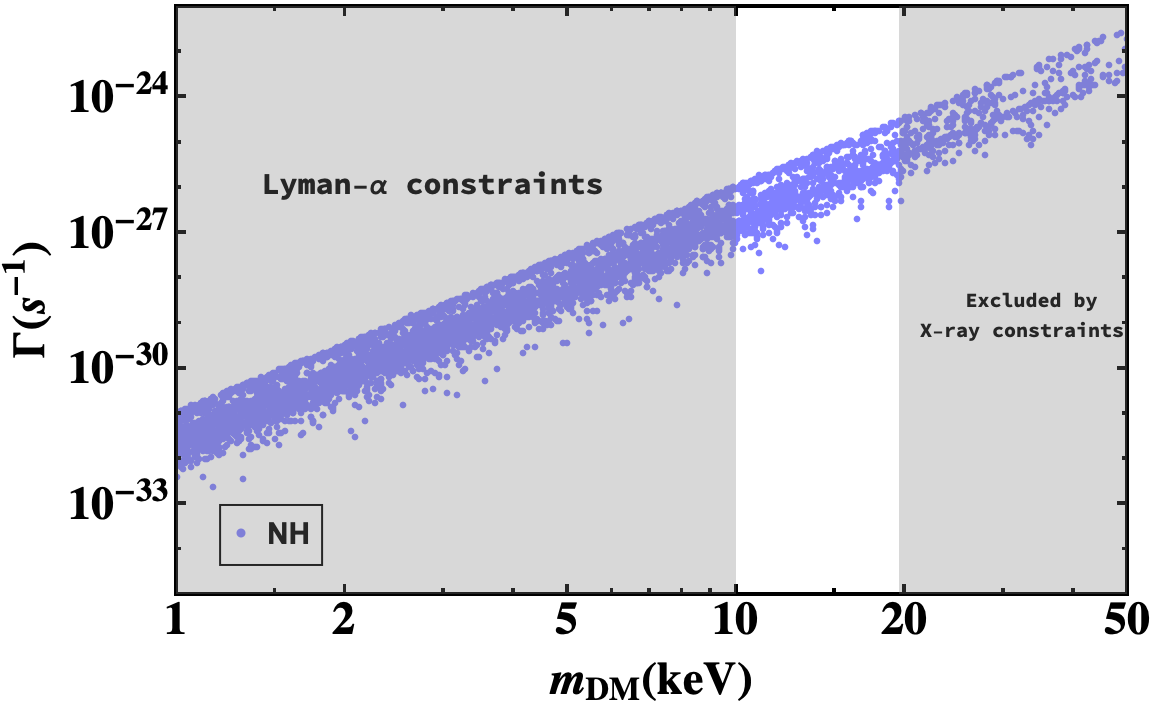}
	\includegraphics[scale=0.4]{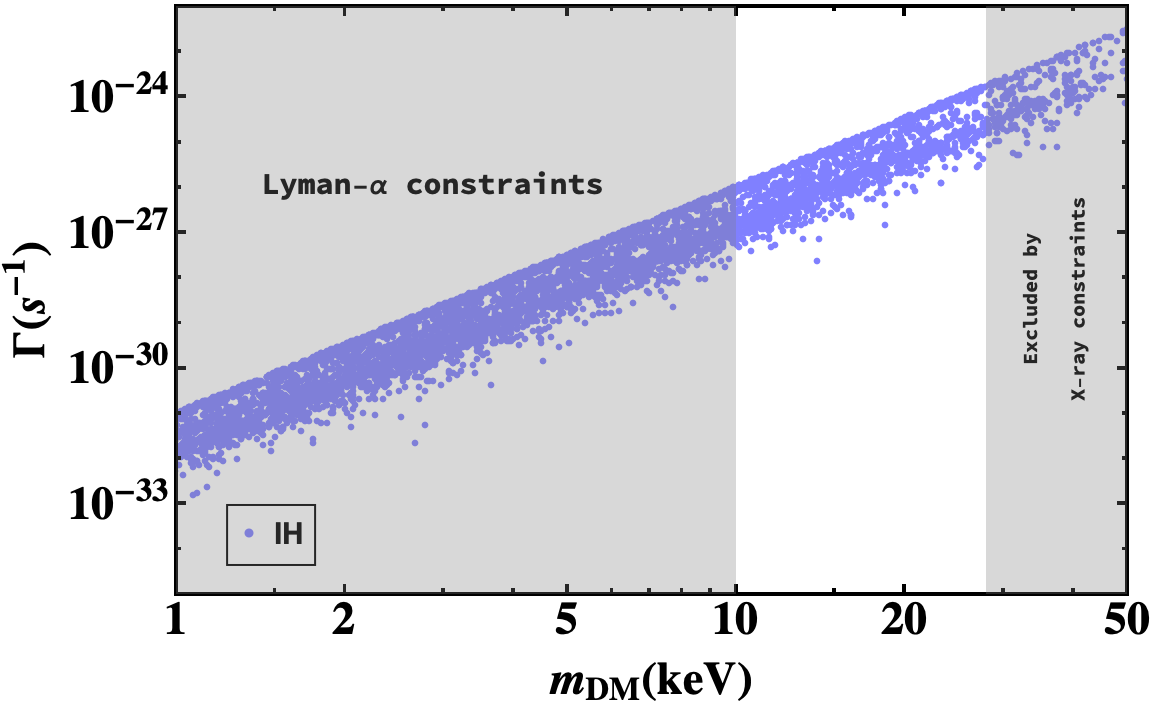}
	\caption{\centering \label{f5}Variation of mass of DM candidate with decay rate for NH(left) and IH(right).}
\end{figure}

\begin{figure}[!h]
	\includegraphics[scale=0.4]{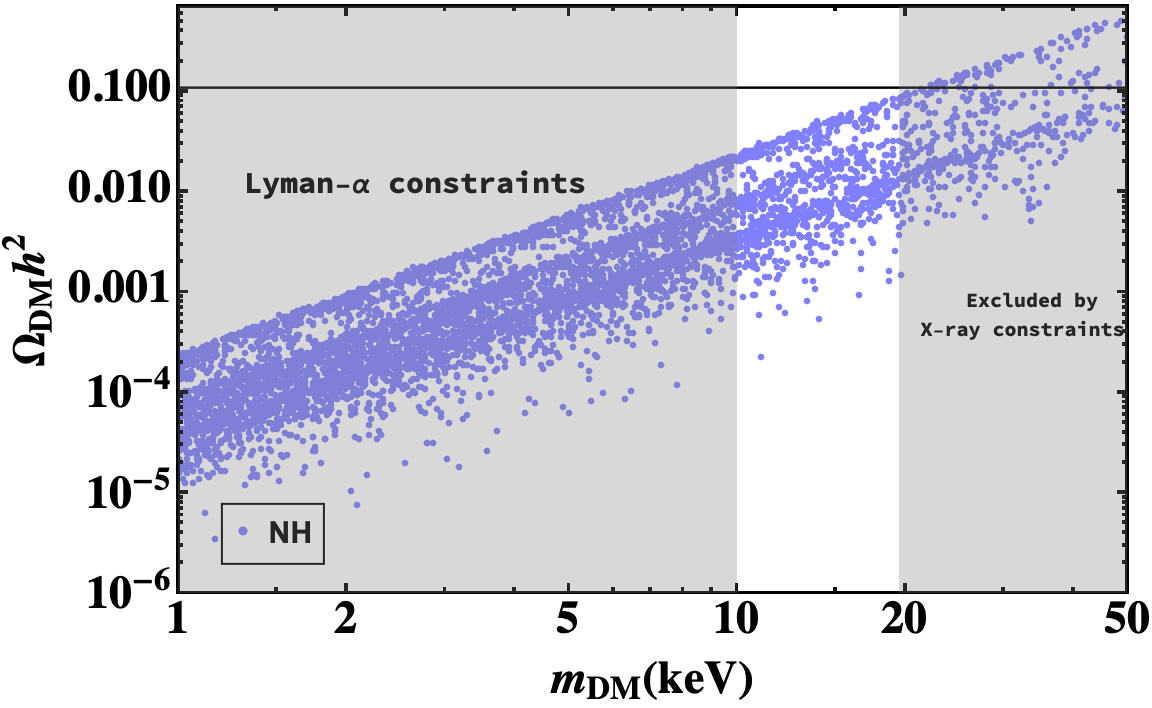}
	\includegraphics[scale=0.4]{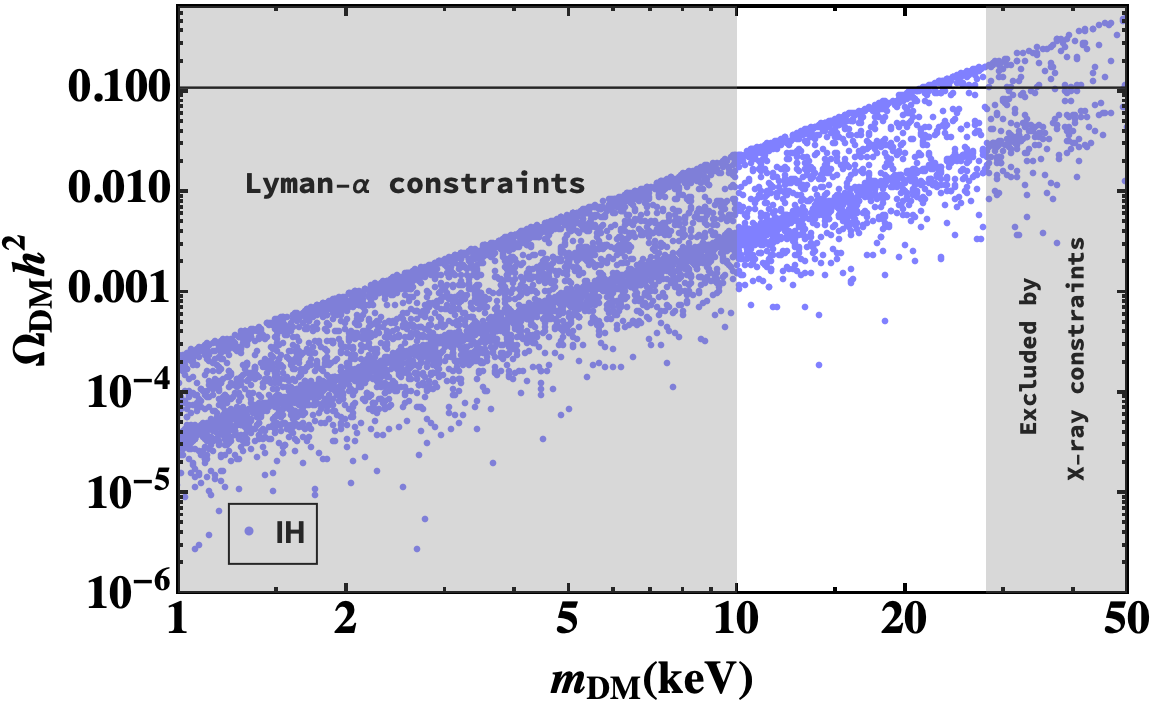}
	\caption{\centering \label{f6}Variation of mass of DM candidate with relic abundance for NH(left) and IH(right).}
\end{figure}

\begin{figure}[!h]
	\includegraphics[scale=0.4]{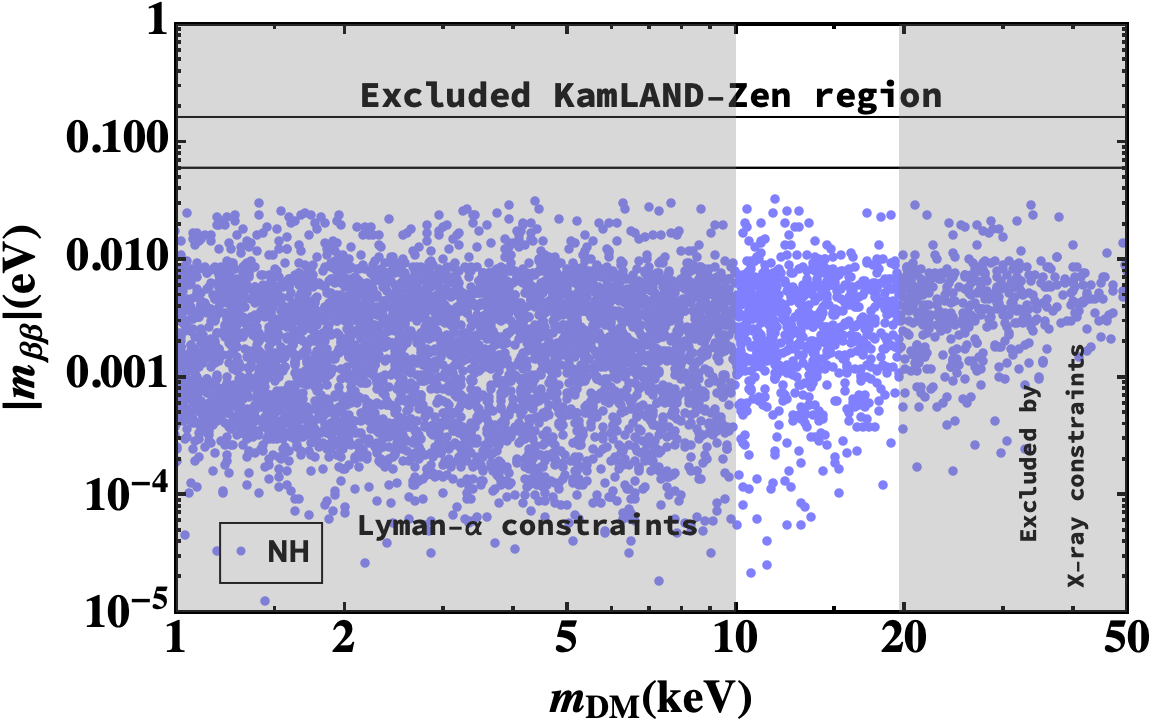}
	\includegraphics[scale=0.4]{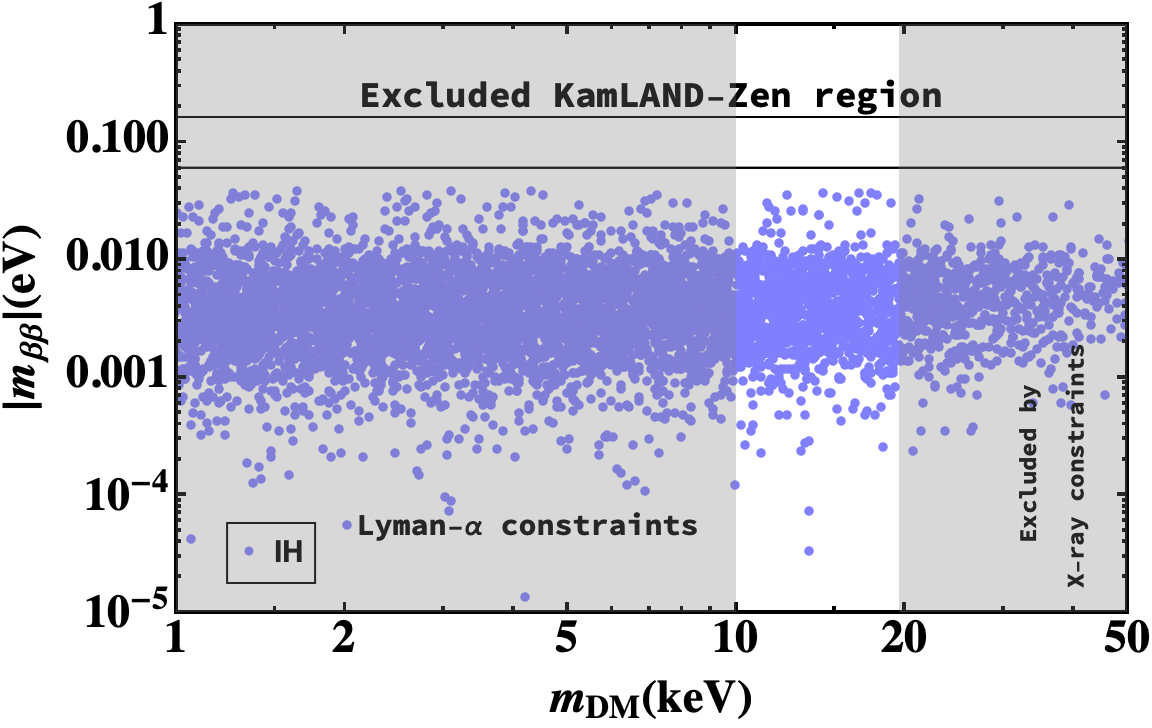}
	\caption{\centering \label{f8}Variation of mass of DM candidate with total effective mass of $0\nu\beta\beta$ for NH(left) and IH(right).}
\end{figure}

\begin{figure}[!h]
	\includegraphics[scale=0.4]{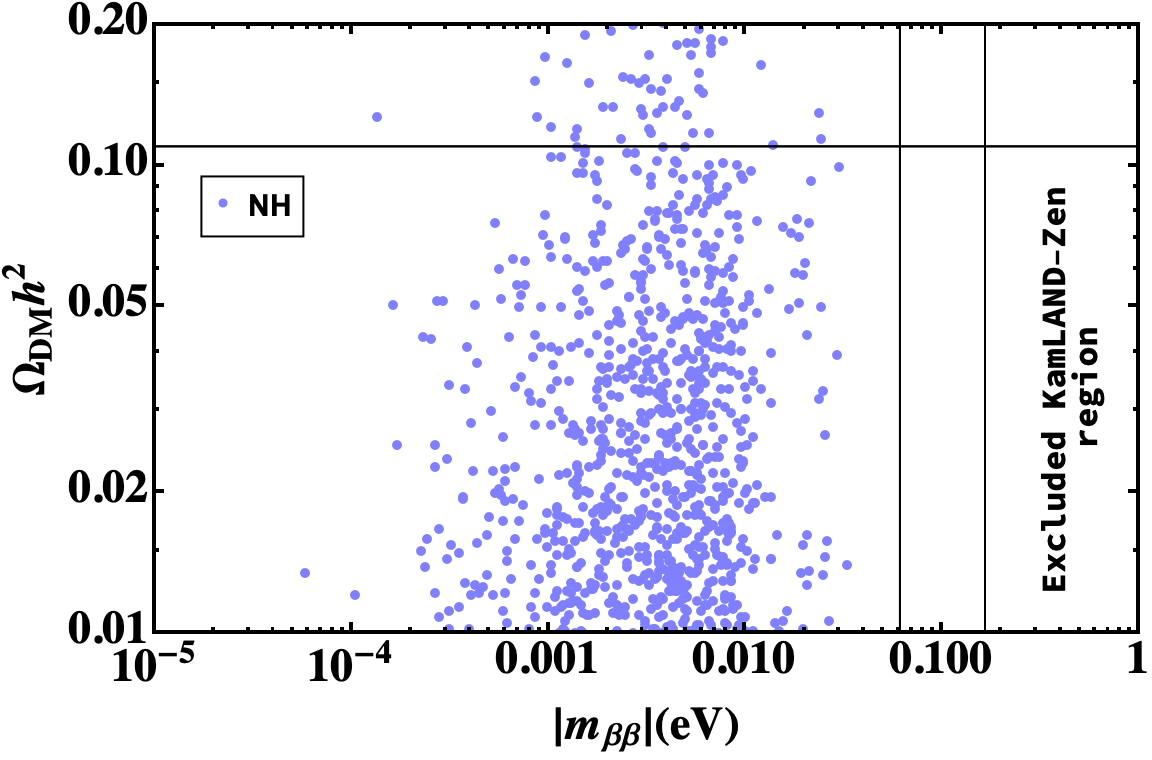}
	\includegraphics[scale=0.4]{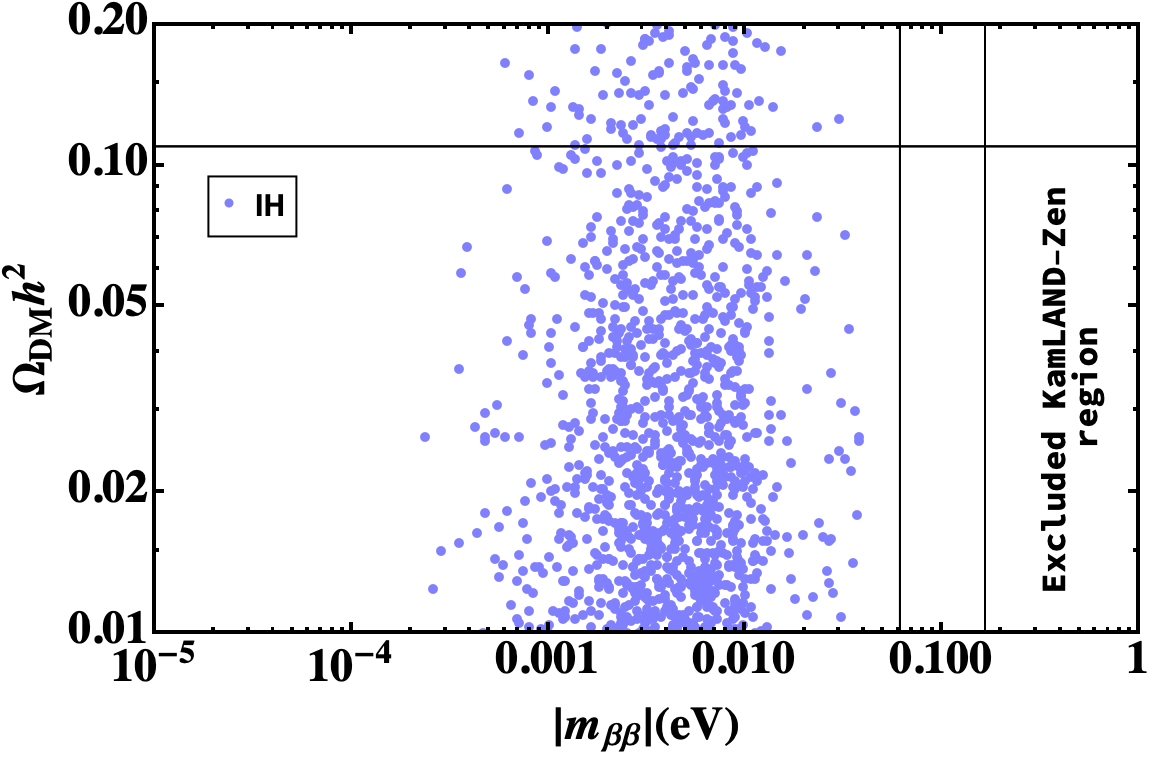}
	\caption{\centering \label{f9}Variation of total effective mass of $0\nu\beta\beta$ with relic abundance for NH(left) and IH(right).}
\end{figure}

\begin{figure}[!h]
	\centering
	\includegraphics[scale=0.4]{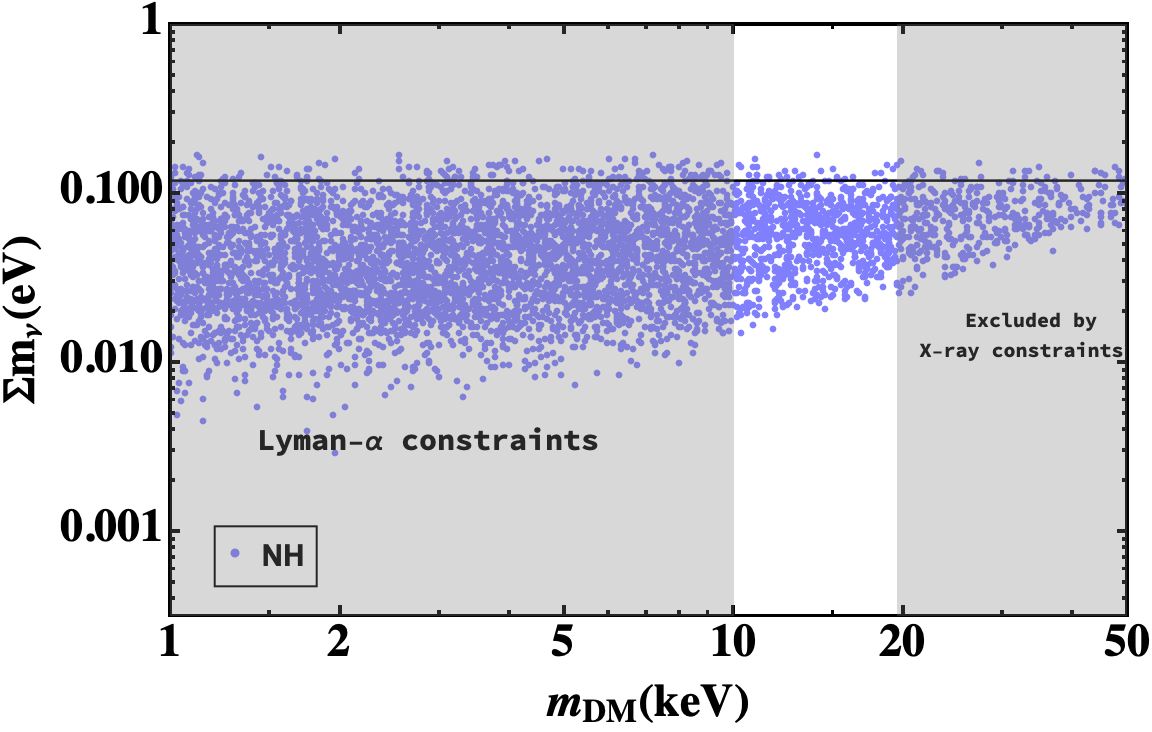}
	\includegraphics[scale=0.4]{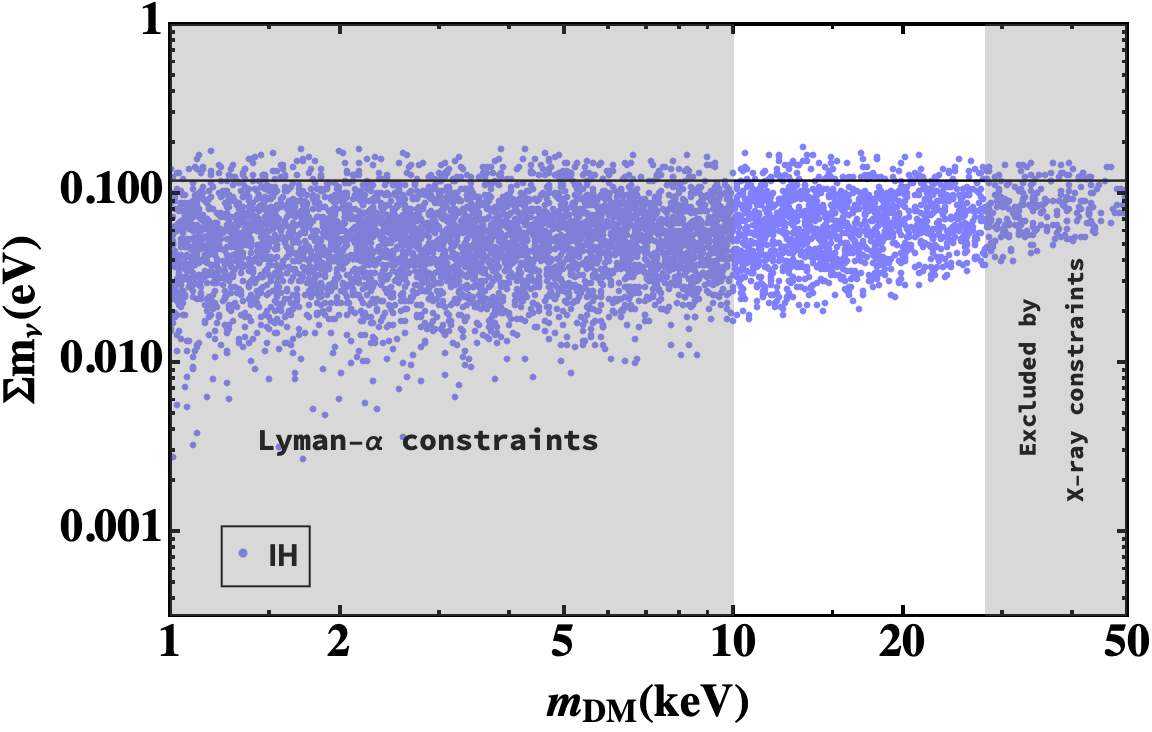}
	\caption{\centering \label{f10}Variation of dark matter mass with sum of neutrino masses for NH(left) and IH(right).}
\end{figure}

\begin{figure}[!h]
	\centering
	\includegraphics[scale=0.4]{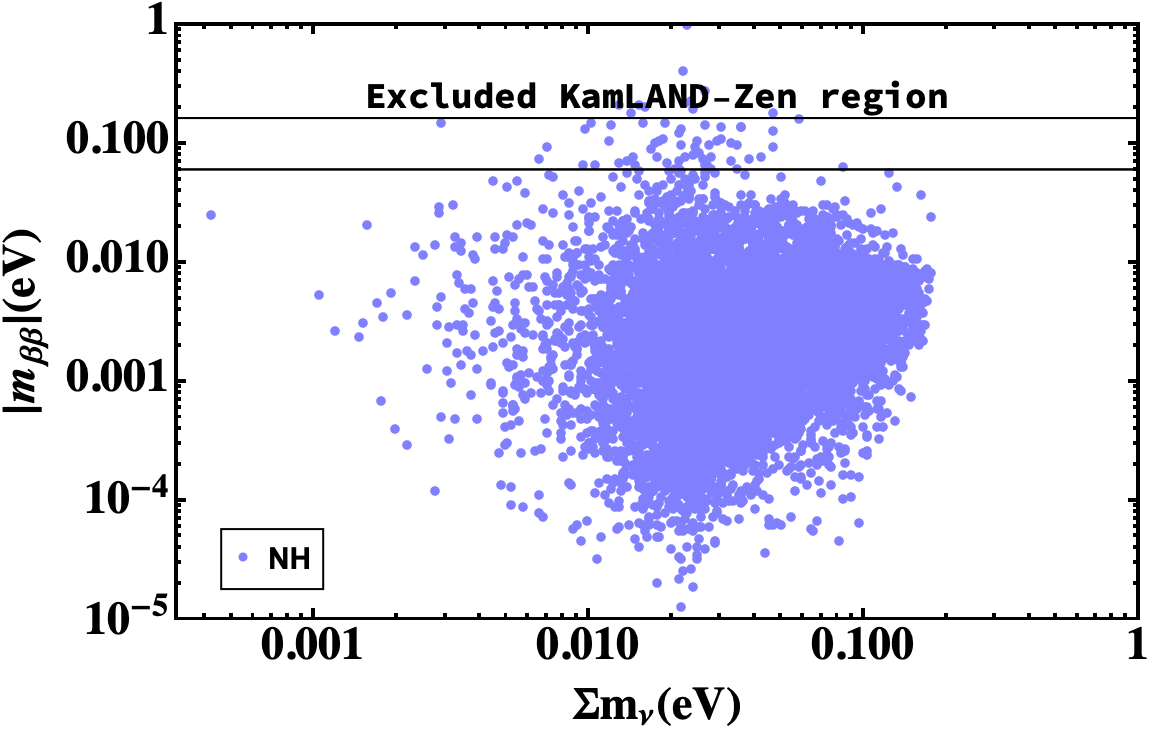}
	\includegraphics[scale=0.4]{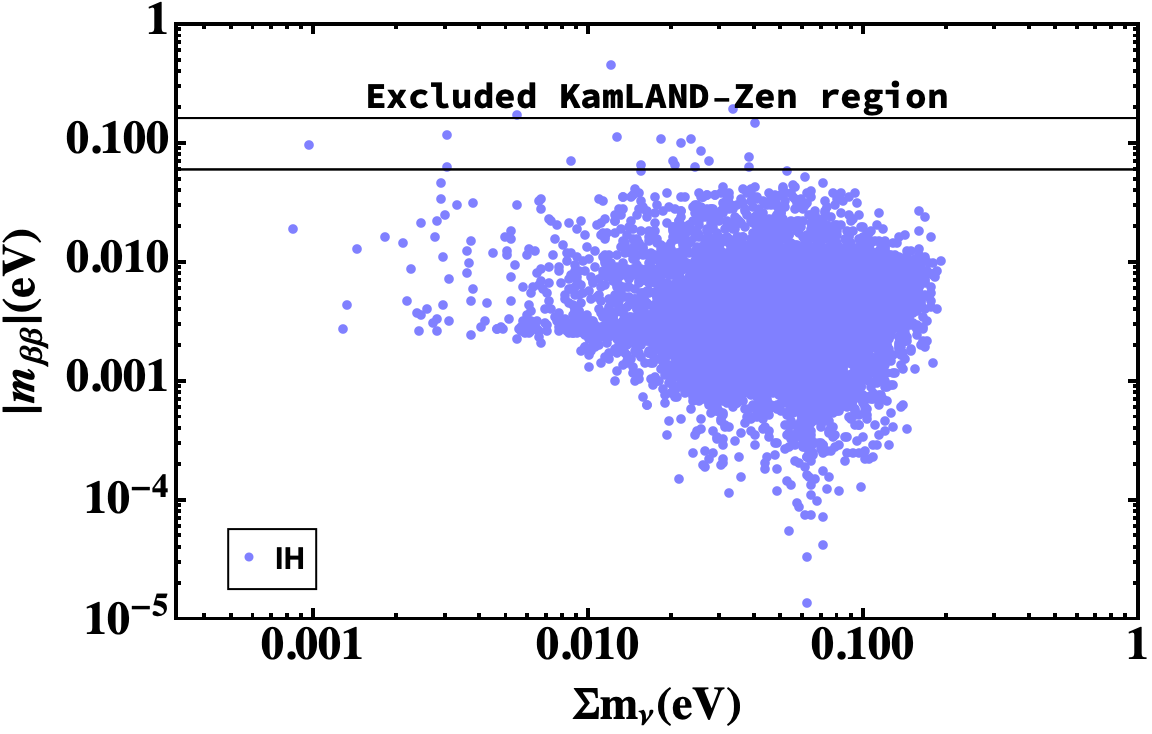}
	\caption{\centering  \label{f11}Variation of effective mass with sum of neutrino masses for NH(left) and IH(right).}
\end{figure}

LRSM has been extended with a sterile fermion per generation as stated in section \ref{lrsm2} and lightest of the same is considered as the DM candidate. In several previous works \cite{Kakoti:2023isn,Kakoti:2023xkn,Boruah:2021ktk}, LRSM has been considered with the scalar sector consisting of one Higgs bidoublet and two scalar triplets and there, the resulting neutrino mass was given as a summation of the type-I and type-II seesaw mass terms. In the current work we consider the scalar sector consisting of one Higgs bidoublet $\phi(1,2,2,0)$ and two scalar doublets $H_{R}(1,1,2,-1)$ and $H_{L}(1,2,1,-1)$. As such, the resulting light neutrino mass is given by the double seesaw mechanism, by the expression in \eqref{e5}. Before determining the resulting neutrino mass, we have calculated the values of the Yukawa couplings $(Y_1,Y_2,Y_3)$ using  equation \eqref{e12a},
\begin{equation}
	\label{e12a}
	M_{\nu}=U_{PMNS}m_{diag}U_{PMNS}^{T}
\end{equation}
where, $U_{PMNS}$ is the Pontecorvo-Maki-Nakagawa-Sakata matrix as described using equation \eqref{e19} and $m_{diag}=diag(m_1,m_2,m_3)$. Once the Yukawa couplings are determined further calculations for phenomenological parameters associated with the mass matrices can be carried out. Here, it is to be noted that for determination of the neutrino parameters we have used the global $3\sigma$ \cite{Esteban:2024eli} values.\\
Since, we are working with $keV$ sterile neutrino as a suitable dark matter candidate, we have taken into consideration the Lyman-$\alpha$ as well as X-ray constraints, and the plots have been done using 'Mathematica'. For our calculations, we have held fixed value of the VEV of the Higgs bidoublet, i.e., $v$ and VEV of the right-handed scalar doublet, $v_{R}$ and $v_{R}=10TeV$. The arbitrary values of $s$ ranges in the order of $1-100$ keV and the dimensionless parameter $f$ has values of the order of $O(10^{-6})$.\\
Now, using equations \eqref{e7} and\eqref{e8}, we calculate the active-sterile mixing, decay rate and relic abundance of the DM candidate and the results have been demonstrated using figures \ref{f4},\ref{f5} and \ref{f6} respectively.
From figure \ref{f4}, it has been observed that for normal hierarchy, the mass of the dark matter candidate ranges from (10-19.6) keV and (10-28) keV for inverted hierarchy following the Lyman-$\alpha$ and X-ray constraints. Figure \ref{f5} shows the variation of mass of dark matter candidate with its decay rate and satisfactorily, the decay rate is found to be negligible.
If we observe figure \ref{f6}, inverted hierarchy shows better result for relic abundance satisfying the observed value of $0.1187$ for mass of the dark matter candidate falling within the allowed range.\\
Coming to $0\nu\beta\beta$ for standard light neutrino contribution, the effective mass of the same is given as,
\begin{equation}
	\label{e18}
	|{m_v^{eff}}|= \sum_{i=1,2,3}{U^2_{ei}}{m_i}
\end{equation}
where, $U_{ei}$ are the elements of the first row of the neutrino mixing matrix $U_{PMNS}$, in which the elements are dependent on known mixing angles $\theta_{13}$ , $\theta_{12}$ and the Majorana phases $\kappa$ and $\eta$. $m_{i}$ are the eigenvalues of the neutrino mass matrix $M_{\nu}$.\\ The $U_{PMNS}$ matrix is given by,
\begin{equation}
	\label{e19}
	U_{PMNS} = \begin{pmatrix}
		c_{12}c_{13} & s_{12}c_{13} & s_{13}e^{-i\delta}\\
		-c_{23}s_{12} - s_{23}s_{13}c_{12}e^{i\delta} & -c_{23}c_{12} - s_{23}s_{12}s_{13}e^{i\delta} & s_{23}c_{13}\\
		s_{23}s_{12} - c_{23}s_{13}c_{12}e^{i\delta} & -s_{23}c_{12} - c_{23}s_{13}s_{12}e^{i\delta} & c_{23}c_{13}
	\end{pmatrix}P
\end{equation}
where, $P = diag(1,e^{i\kappa}, e^{i\eta})$. So the effective mass can be parametrized in terms of the elements of the diagonalizing matrix and the eigenvalues as,
\begin{equation}
	\label{e20}
	{m_v^{eff}} = m_1c_{12}^{2}c_{13}^{2} + m_2s_{12}^{2}c_{13}^{2}{e^{2i\kappa}} + m_3s_{13}^{2}{e^{2i\eta}}.
\end{equation}
For heavy right-handed neutrino contribution the effective mass is given by,
\begin{equation}
	\label{e21}
	|m_N^{eff}| = <p^{2}> \sum_{j=1,2,3}\frac{({V^{\nu_{R}\nu_{R}}}_{ej})^{2}}{M_{N_{j}}}
\end{equation}
and, for sterile neutrino contribution, the effective mass is given by 
\begin{equation}
	\label{e22}
	|m_S^{eff}| = <p^{2}> \sum_{k=1,2,3}\frac{({V^{\nu_{R}S}_{ek}})^{2}}{M_{S_{k}}}
\end{equation}
where, $p^{2}$ is the typical momentum exchange of the process.
So, the total contribution coming from all the above mentioned processes can be given by,
\begin{equation}
	\label{e23}
	|m_{total}^{eff}| = \sqrt{( \sum_{i=1,2,3}{U^2_{Li}}{m_i})^{2}+<p^{2}>^{2}\Bigg(\sum_{j=1,2,3}\frac{({V^{\nu_{R}\nu_{R}}}_{ej})^{2}}{M_{N_{j}}}+\sum_{k=1,2,3}\frac{({V^{\nu_{R}S}_{ek}})^{2}}{M_{S_{k}}}\Bigg)^{2}}
\end{equation}
where, $M_{N_{j}}$ and $M_{S_{k}}$ are the mass eigenvalues for heavy right-handed neutrinos and sterile neutrino respectively. In the framework under consideration, $U=U_{PMNS}$, $V^{\nu_{R}\nu_{R}} \approx U_{N}$ which is the mixing matrix for heavy right-handed neutrinos and $V^{\nu_{R}S} \equiv M \mu^{-1}U_{S}$, where $U_{S}$ is the mixing matrix for sterile neutrinos.
We have calculated the total contribution using equation \eqref{e23} and analysis have been carried out to check if $0\nu\beta\beta$ is possible for the allowed mass range of the DM candidate and the results are shown in figures \ref{f8} and \ref{f9}.
Finally, we have also plotted the variation of phenomenological parameters with sum of neutrino masses to check the consistency of the model. The results are shown in figures \ref{f10} to \ref{f11}.
\section{\label{lrsm6}Discussion}

As described in the preceding sections, we have extended LRSM with a sterile fermion per generation and considered the lightest of the same as the suitable dark matter candidate. Realization of the model has been done using modular group $\Gamma_3$ with weight 2. As it is a matter of fact that modular groups are not finite and as such, we consider the condition that for level of the modular group $N$,  for $2\leq N \leq 5$, the particular modular group is isomorphic to a non-abelian discrete symmetry group which is finite and in that case, we consider the properties of that particular discrete symmetry group for the realization of the model.\\
$\Gamma_3$ modular group of weight 2 is isomorphic to $A_4$ discrete symmetry group, as such we consider the $A_4$ multiplication rules for the determination of Dirac, Majorana, active-sterile mixing as well as sterile neutrino mass matrices. After determination of the respective mass matrices given by equations \eqref{e4},\eqref{e:a} and \eqref{e:b}, the resulting neutrino mass matrix is determined as shown in \eqref{e5}. Now that the neutrino mass matrix is determined, we calculate the values of $Y_1$, $Y_2$ and $Y_3$ we make use of the expression \eqref{e12a}.
For calculation of the parameters, we have used the global $3\sigma$ values. Now that, all the parameters which forms the resulting light neutrino mass matrix are known, we can go on for the further calculations for analysis of dark matter and phenomenological study including $0\nu\beta\beta$. \\
\begin{itemize}
	\item The principal objective of this work lies on the extension of LRSM with a sterile fermion per generation so as to incorporate the study of dark matter within the model. The lightest of sterile fermion is considered as the candidate for dark matter and the light neutrino mass in the current study is generated by the double seesaw mechanism. For phenomenological study, neutrinoless double beta decay $(0\nu\beta\beta)$ has been taken into consideration. The points below clearly state the results obtained in the current work.
	
	\item Firstly, we calculate the active-sterile mixing angle which helps in determining the mass range of dark matter candidate satisfying both Lyman $\alpha$ as well as X-ray constraints. For a keV sterile neutrino to be a good dark matter candidate, its mass should normally lie in the range of 0.4-50 keV and the mixing angle typically lies within $10^{-12}-10^{-8}$. Figure \ref{f4} clearly depicts that for normal hierarchy, mass of dark matter lies within $10-19.6$ keV and for inverted hierarchy it ranges from 10 to 28 keV. The mixing angle for both hierarchies have been determined to lie within $10^{-11}-10^{-9}$ which is a satisfactory result.
	\item Relic density and decay rate of the dark matter candidate have been calculated using equations \eqref{e7} and \eqref{e8}. Figure \ref{f5} shows the variation of mass of DM candidate with its decay rate. Although the decay rate ranges from around $10^{-33}$ to around $10^{-23}$, but for dark matter mass satisfying both the constraints, the decay rate falls within $10^{-27}-10^{-24}$.
	\item From the variation of relic abundance with dark matter  mass  shown in figure \ref{f6}, it can be inferred that results are better for inverted hierarchy as the data points pertaining to the observed value of relic abundance falls within the allowed range of dark matter mass, unlike the case of normal hierarchy.
	\item We have tried to analyze the impact of double seesaw mechanism on $0\nu\beta\beta$ along with the study of dark matter within the context of the present work. Figures \ref{f8} and \ref{f9} depict the variation of total effective mass corresponding to standard light neutrino, heavy right-handed and sterile neutrino contribution with dark matter mass and relic abundance  respectively. From the figures, it can be inferred that the data obtained satisfy the constraints for dark matter mass and experimental bound for $0\nu\beta\beta$ simultaneously. However, it has been observed that very less data point satisfy the observed value of relic abundance when within the experimental bounds of $0\nu\beta\beta$ for both normal and inverted hierarchy.
	\item Towards the end of numerical analysis and results, we have shown the variations corresponding to sum of neutrino masses with various phenomenological parameters and the results obtained have been depicted in figures \ref{f10} to \ref{f11}.
\end{itemize}
\section{\label{lrsm7}Conclusion}
Left-Right Symmetric Model has been successful in explaining the generation of neutrino masses and mixing as well as the associated phenomenology. However, our prime focus is to determine how fruitful is the incorporation of modular symmtery within the context of the model. And in the current work, we have emphasized on the study and analysis of dark matter within the context of modular symmetric LRSM. A brief summary of the work has been presented as under.
\begin{itemize}
	\item In the present work, LRSM has been extended with a sterile fermion  per generation and the light neutrino mass is determined using the double seesaw framework. The scalar sector in the model consists of a Higgs bidoublet $\phi(1,2,2,0)$, two scalar doublets $H_R(1,1,2,-1)$ and $H_L(1,2,1,-1)$. Previous works like \cite{Kakoti:2023isn},\cite{Kakoti:2023xkn} have also taken into consideration the triplets that is, $\Delta_R(1,1,3,2)$ and $\Delta_L(1,3,1,2)$ in the scalar sector.
	\item The model has been realized using $A_4$ modular symmetry. We can also use other symmetry groups depending upon the level of modular form used and as such, the number of modular forms will also change, as per the table \ref{table:3} in the manuscript. As we have used $A_4$ modular symmetry, there are three modular forms $(Y_1,Y_2,Y_3)$ and hence it became easier for us to construct the $3 \times 3$ mass matrices in terms of the aforementioned modular forms which are consistent with the neutrino data. 
	\item The Yukawa couplings as calculated using equation \eqref{e12a} are found to satisfy the condition that its modulus is greater than unity.
	\item Talking about dark matter which is the primary aspect of the current work, the lightest of the sterile fermion added per generation can indeed be recognized as a suitable dark matter candidate for the model under consideration.
	\item  The total effective mass for $0\nu\beta\beta$ corresponding to standard light neutrino, heavy right-handed and sterile neutrino contribution have been calculated and less number of  points in the scattered plots are found to satisfy the experimental constraints, for both normal and inverted hierarchies.
	\item However, as stated in the introduction, as lepton flavor violation in doublet LRSM can take place in loop levels, the study of the same has been kept for future work.
\end{itemize}
From the above discussions, it can be stated that taking into consideration the extension of LRSM with the sterile fermion, dark matter and $0\nu\beta\beta$ can be studied simultaneously as the effective mass as well as mass of the DM candidate and relic abundance are all found to satisfy the experimental bounds, Lyman-$\alpha$ constraints and the observed value of relic abundance respectively.\\
Use of modular symmetry can also be stated to be fruitful and advantageous, owing to non-requirement for use of extra particles within the context of the model. Also, results obtained as discussed above, are found to satisfy the experimental constraints. As such, conclusively it can be brought into attention that different phenomenology in Beyond Standard Model framework can be studied using modular symmetry as it can satisfactorily explain them and at the same time keeps the model minimal.

\begin{center}
	\section*{Appendix A : Properties of $A_4$ discrete symmetry group.}
\end{center}

$A_4$ is a non-abelian discrete symmetry group which represents even permutations of four objects. It has four irreducible representations, three out of which are singlets $(1,1',1'')$ and one triplet $3$ ($3_A$ represents the anti-symmetric part and $3_S$ the symmetric part). Products of the singlets and triplets are given by,
\begin{center}
	\begin{equation*}
		1 \otimes 1 = 1
	\end{equation*}
\end{center}
\begin{center}
	\begin{equation*}
		1' \otimes 1' = 1''
	\end{equation*}
\end{center}
\begin{center}
	\begin{equation*}
		1' \otimes 1'' = 1
	\end{equation*}
\end{center}
\begin{center}
	\begin{equation*}
		1'' \otimes 1'' = 1'
	\end{equation*}
\end{center}
\begin{center}
	\begin{equation*}
		3 \otimes 3 = 1 \oplus 1' \oplus 1'' \oplus 3_A \oplus 3_S
	\end{equation*}
\end{center}
If we have two triplets under $A_4$ say, $(a_1,a_2,a_3)$ and $(b_1,b_2,b_3)$, then their multiplication rules are given by,
\begin{center}
	\begin{equation*}
		1 \approx a_1b_1 + a_2b_3 + a_3b_2
	\end{equation*}
	\begin{equation*}
		1' \approx a_3b_3 + a_1b_2 + a_2b_1
	\end{equation*}
	\begin{equation*}
		1'' \approx a_2b_2 + a_3b_1 + a_1b_3
	\end{equation*}
	\begin{equation*}
		3_S \approx \begin{pmatrix}
			2a_1b_1-a_2b_3-a_3b_2 \\
			2a_3b_3-a_1b_2-a_2b_1 \\
			2a_2b_2-a_1b_3-a_3b_1
		\end{pmatrix}
	\end{equation*}
	\begin{equation*}
		3_A \approx \begin{pmatrix}
			a_2b_3-a_3b_2 \\
			a_1b_2-a_2b_1 \\
			a_3b_1-a_1b_3
		\end{pmatrix}
	\end{equation*}
\end{center}

\begin{center}
	\section*{Appendix B : Modular Symmetry}
\end{center}

Modular symmetry has gained much importance in aspects of model building \cite{King:2020qaj}, \cite{Novichkov:2019sqv}. This is because it can minimize the extra particle called 'flavons' while analyzing a model with respect to a particular symmetry group. An element $q$ of the modular group acts on a complex variable $\tau$ which belongs to the upper-half of the complex plane given as \cite{Novichkov:2019sqv} \cite{Feruglio:2017spp} 
\begin{equation*}
	\label{E:23}
	q\tau = \frac{a\tau+b}{c\tau+d}
\end{equation*}
where $a,b,c,d$ are integers and $ad-bc=1$, Im$\tau$$>$0.\\
The modular group is isomorphic to the projective special linear group PSL(2,Z) = SL(2,Z)/$Z_2$ where, SL(2,Z) is the special linear group of integer $2\times2$ matrices having determinant unity and $Z_2=({I,-I})$ is the centre, $I$ being the identity element. The modular group can be represented in terms of two generators $S$ and $T$ which satisfies $S^2=(ST)^3=I$. $S$ and $T$ satisfies the following matrix representations:
\begin{equation*}
	\label{E:24}
	S = \begin{pmatrix}
		0 & 1\\
		-1 & 0
	\end{pmatrix}
\end{equation*}
\begin{equation*}
	\label{e:25}
	T = \begin{pmatrix}
		1 & 1\\
		0 & 1
	\end{pmatrix}
\end{equation*}
corresponding to the transformations,
\begin{equation*}
	\label{e:26}
	S : \tau \rightarrow -\frac{1}{\tau} ; T : \tau \rightarrow \tau + 1
\end{equation*}
Finite modular groups (N $\leq$ 5) are isomorphic to non-abelian discrete groups, for example, $\Gamma_{3} \approx A_4$, $\Gamma_{2} \approx S_3$, $\Gamma_{4} \approx S_4$. While using modular symmetry, the Yukawa couplings can be expressed in terms of modular forms, and the number of modular forms present depends upon the level and weight of the modular form. For a modular form of level $N$ and weight 2k, the table below shows the number of modular forms associated within and the non-abelian discrete symmetry group to which it is isomorphic \cite{Feruglio:2017spp}.
\begin{table}[!h]
	\begin{center}
		\begin{tabular}{|c|c|c|}
			\hline
			N & No. of modular forms & $\Gamma_{N}$ \\
			\hline
			2 & k + 1 & $S_3$ \\
			\hline
			3 & 2k + 1 & $A_4$ \\
			\hline
			4 & 4k + 1 & $S_4$ \\
			\hline
			5 & 10k + 1 & $A_5$ \\
			\hline 
			6 & 12k &  \\
			\hline
			7 & 28k - 2 & \\
			\hline
		\end{tabular}
		\caption{\label{table:3}No. of modular forms corresponding to modular weight 2k.}
	\end{center}
\end{table}
In our work, we have used modular form of level 3, that is, $\Gamma_3$ which is isomorphic to $A_4$ discrete symmetry group. The weight of the modular form is taken to be 2, and hence it will have three modular forms $(Y_1,Y_2,Y_3)$ which can be expressed as expansions of q given by,
\begin{equation*}
	\label{e:27}
	Y_1 = 1 + 12 q + 36 q^2 + 12 q^3 + 84 q^4 + 72 q^5 + 36 q^6 + 96 q^7 + 
	180 q^8 + 12 q^9 + 216 q^{10}
\end{equation*}
\begin{equation*}
	\label{e:28}
	Y_2 = -6 q^{1/3} (1 + 7 q + 8 q^2 + 18 q^3 + 14 q^4 + 31 q^5 + 20 q^6 + 
	36 q^7 + 31 q^8 + 56 q^9)
\end{equation*}
\begin{equation*}
	\label{e:29}
	Y_3 = -18 q^{2/3} (1 + 2 q + 5 q^2 + 4 q^3 + 8 q^4 + 6 q^5 + 14 q^6 + 
	8 q^7 + 14 q^8 + 10 q^9)
\end{equation*}
where, $q = \exp(2\pi i \tau)$.

\section*{Acknowledgments}
Ankita Kakoti acknowledges Department of Science and Technology (DST), India (grant DST/INSPIRE Fellowship/2019/IF190900) for the financial assistantship.
\bibliography{cite3a}
\bibliographystyle{unsrt}
\end{document}